\documentclass[aps,pre,twocolumn,floatfix,showpacs,amsmath,amssymb]{revtex4}

\usepackage{graphicx}% Include figure files
\usepackage{bm}% bold math
\usepackage{amssymb}% bold math

\begin{document}

\newcommand{\bsigma}{\bm{\sigma}}
\newcommand{\VIS}{\textit{VIS}}
\newcommand{\sigmaY}{\sigma_{\text{Y}}}

\title{Internal Stress in a Model Elasto-Plastic Fluid}

\author{\surname{Ooshida} Takeshi}  
\email{ooshida@damp.tottori-u.ac.jp}

\affiliation{%
  Department of Applied Mathematics and Physics,
  Tottori University, JP-680-8552, Japan
}

\author{Ken Sekimoto}\email{sekimoto@turner.pct.espci.fr}

\affiliation{MSC - CNRS,  Universit\'e Paris 7, 
2 place Jussieu, F-75251 Paris, France}\thanks{Also at: 
Physico-Chimie Th\'eorique, E.S.P.C.I., 10 rue Vauquelin, 
F-75231 Paris, France}

\date{\today}

\begin{abstract}
Plastic materials can carry  memory of 
past mechanical treatment
in the form of internal stress.
We introduce a natural definition of the vorticity of 
internal stress
in a simple two-dimensional model of elasto-plastic fluids,
which generates the internal stress.
We demonstrate how the internal stress is induced
under external loading, and how the presence of
the internal stress modifies the plastic behavior.
\end{abstract}

\pacs{83.60.La, 46.35.+z, 62.20.Fe, 61.20.Lc}% PACS

\maketitle

%%%%%%%%%%%%%%%%%%%%%%%%%%%%%%%%%%%%%%%%%%%%%%%%%%%%%%%%%%

The \textrm{internal stress}, also
known as the remanent stress in material science, 
is the stress which is maintained in a material 
without external mechanical supports.
The internal stress is found in both
discrete and continuous systems with widely 
different spatial scales.
An examples of macroscale is the {``{tensegrity}''} of 
architecture~\cite{architecture}, where the internal
balance of the structural elements under tension
and those under compression maintains the solidity of 
the architecture. Its putative counterpart in biological 
cells has also been proposed~\cite{tensegrity}, where the 
intracellular cytoskeletal networks are claimed to
play the role of the structural elements.
In lattice mechanics of solids, we could mention as 
examples of internal stress the lattice mismatch 
of crystal growth~\cite{lattice_mismatch}, the 
dislocations~\cite{Landau,Friedel,Eshelby},
the phase separation in alloys into 
the phases of different lattice constants~\cite{Cahn}.
In polymers, rubbers vulcanized under more than one 
deformed states (called the permanent set by 
Flory~\cite{Flory}) would be an example of internal stress.
The stresses in those systems, either in atomic or 
mesoscopic scales, are distributed and establish 
the mechanical balance among them.
The internal stress is thus a very common phenomenon.

%%%%%%%%%%%%%%%%%%%%%%%%%%%%%%%%%%%%%%%%%%%%%%%%%%%%%%%%%%%
As the internal stress is not describable by the 
monovalent and continuous elastic displacements~\cite{Landau,Friedel},
Eshelby had introduced the ``incompatibility tensor''~\cite{Eshelby}
in the context of full three-dimensional lattice defects.
Despite the several methods developed so far to describe 
the internal stress, few studies has been done on the
\emph{dynamics} of the internal stress associated with plastic flows.
Also little has been studied on the hysteresis related to the 
internal stress, especially on the plastic yielding under 
macroscopic deformations of the system including internal
stress~\cite{Miyamoto,Nakahara}.
In this {\it Letter} we try to shed light on these problems 
through a simple model of elasto-plastic 
fluid in which we may use the analogue of
the dislocation density~\cite{Landau,Friedel}
to characterize 
the internal stress.
Our model is basically the Bingham model of plastic 
fluid~\cite{Bingham} supplemented with a finite 
shear compliance. This is a version of generalized 
Maxwell models of viscoelastic fluid with a longtime 
relaxation~\cite{derec,Miyamoto}.
After describing our model, 
we will show how an \textrm{internal stress} is dynamically
brought into a system under the external loading, and how
the internal stress is maintained after the removal of the 
loading.
Although the local threshold stress  of plastic yielding 
(called the yield stress) is constant everywhere, 
the internal stress maintained turns out to be generally 
subthreshold and inhomogeneous.
This implies that the system with yield stress can bear
the memory of past operations through the internal stress.
Moreover, we will show that the presence of such internal 
stress modifies the plastic response of the system
in a rather intriguing manner.
%%%%%%%%%%%%%%%%%%%%%%%%%%%%%%%%%%%%%%%%%%%%
\begin{figure}[bt]
 \includegraphics[width=8.5cm,clip]{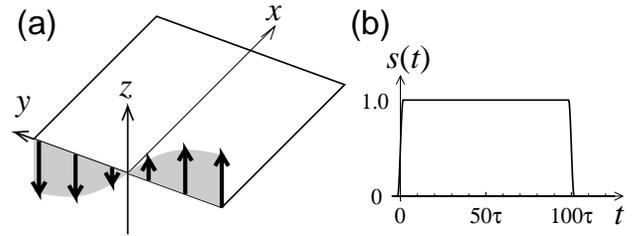}
 \caption{%
(a) System's geometry and the initial transient loading 
 on the boundary (thick arrows) at $x=0$. 
The fluid velocity $\partial{u(x,y,t)}/\partial t$ 
lies along the $z$-axis.
(b) Time protocol of loading, $s(t)$ (see the text).}
 \label{fig:setup}
\end{figure}
%%%%%%%%%%%%%%%%%%%%%%%%%%%%%%%%%%%%%%%%%%%%

%%%%%%%%%%%%%%%%%%%%%%%%%%%%%%%%%%%%%%%%%%%%%%%%%%%%%%%%%%%
{Our model is constructed as follows.}
We constrain the fluid to flow in $\|z$ direction
while its flow profile may have heterogeneities 
in the lateral ($x,y$) directions.
We denote the displacement of the fluid along the 
{$z$-axis} by $u= u(x,y,t)$.
The shear stress components are then essentially 
represented by the two-dimensional vector, 
$\bsigma=(\sigma_{xz}(x,y,t),
\sigma_{yz}(x,y,t)).$ 
The evolution of $u$ and $\bsigma$ is
given by the following isotropic continuum model:
\begin{equation}\label{eq:newton2}
  \rho \frac{{\partial}^2 u}{{\partial}t^2} 
  = \nabla\cdot \bsigma + F^{\text{ext}},
\end{equation}
\begin{equation}\label{eq:elpl}
  \frac{1}{\mu}\frac{{\partial}{\bsigma}}{{\partial}t}
  +\frac{1}{\eta} \bm{\Phi}(\bsigma)=\nabla
  \frac{{\partial}u}{{\partial}t},
\end{equation}
where
$\nabla \equiv (\partial/\partial x,\partial/\partial y)$ and
$F^{\text{ext}}$ is the external force density,
which we incorporate only as a boundary loading (see below).
$\rho$ and $\mu$ are, respectively,
the mass density and the elastic modulus
 per unit length along the $z$-axis. 
$\bm{\Phi}(\bm{\sigma})$ takes a two-dimensional vector value
which is parallel to $\bm{\sigma}$.
This can be regarded
as one of the generalized Maxwell models~\cite{derec,Miyamoto},
whose original form is recovered by replacing 
$\bm{\Phi}(\bsigma)$ by $\bsigma$.
The special feature ascribed to the Bingham model~\cite{Bingham} 
is that $\bm{\Phi}(\bsigma)$  includes  a threshold (yield) stress, 
$\sigmaY$, of the plastic yielding: 
\begin{equation}\label{eq:wpl2}
\bm{\Phi}(\bsigma)
 = \begin{cases}
    0 &  \text{(for ${|\bsigma|}<{\sigmaY}$)} \\
    \bsigma-\dfrac{\bsigma}{|\bsigma |} \sigmaY  
    &    \text{(for ${|\bsigma|}>{\sigmaY}$)}
   \end{cases}.
\end{equation}
The parameter $\eta$ is the viscosity coefficient per 
unit length along the $z$-axis.
Although not shown, this model can reproduce
the one-dimensional propagation of the yielding
(i.e.~fluidization) front at the speed of
the  transverse elastic wave,
$ {v_{\text{s}}} = \sqrt{\mu/\rho}$.
This process of fluidization, which is consistent with a
numerical observation~\cite{Varnik}, is not describable by the
classical { ``Bingham model''} where $\mu=\infty$.

Much more spectacular is the behavior in the higher
dimensionalities, which can support the \textrm{internal stress}.
In our model, we define 
  the \emph{vorticity of the internal stress} ({\VIS}  for short),  
which we denote by $\omega$, as follows:
\begin{equation} \label{eq:IScont}%
\omega \equiv
\frac{\partial
\sigma_{yz}}{\partial x} - %
\frac{\partial \sigma_{xz}}{\partial y}.%
\end{equation}
The {\VIS} corresponds to the dislocation density in 
the continuum description of lattice defects~\cite{Landau,Friedel},
expressed in terms of stress.
Our model  (\ref{eq:elpl}) gives the evolution of the \VIS:
\begin{equation} \label{eq:GS}%
 \frac{\partial \omega}{\partial t}  \, %
 = - \mathrm{div}\bm{J},\quad
\bm{J}\equiv \frac{\mu}{\eta}
\left( \begin{matrix}  {\Phi_y} \cr {-\Phi_x} \end{matrix}\right).
\end{equation}
The first equation indicates that $\bm{J}$ is 
the conservative flux of $\omega$. 
Eq.~(\ref{eq:GS}) implies that $\omega$ is constant in time
where there is no yielding, $\bm{\Phi}=0$. It also implies that
the integral of {\VIS} over a region is invariant if this 
 is surrounded by a non-yielding region.  
%
%%%%%%%%%%%%%%%%%%%%%%%%%%%%%%%%%%%%%%%%%%%%
\begin{figure}[bt]
 \includegraphics[width=8.4cm] {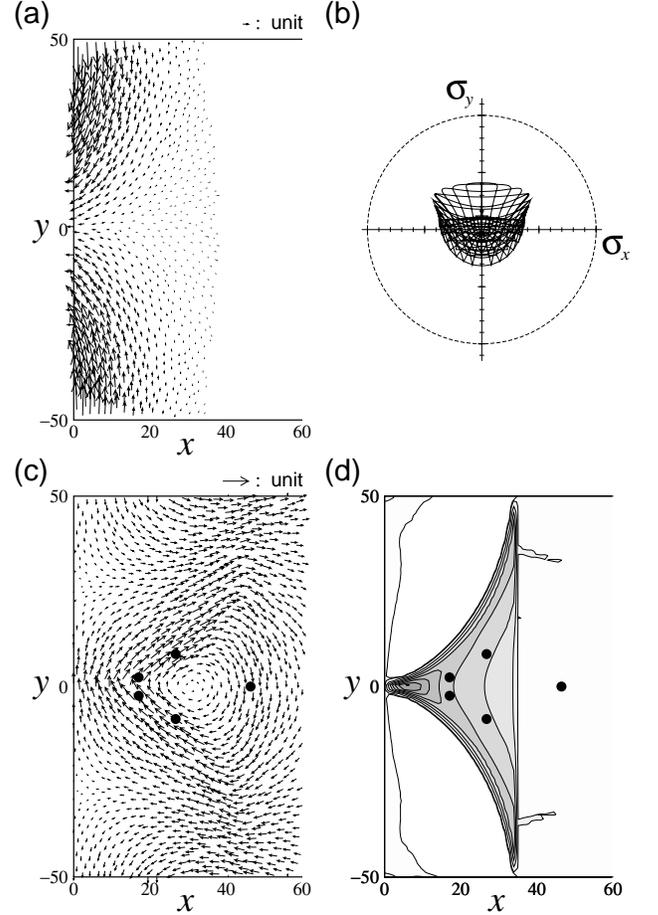}
\caption{%
(a): The flux of the vorticity of the internal stress 
(\VIS), $\protect\bm{J}$ in units of $\sigmaY/\tau$, at $t=100\tau$. 
(b): The spatially distributed 
stress components $(\sigma_{xz},\sigma_{yz})$ at $t=500\tau$ 
shown in the $(\sigma_{xz},\sigma_{yz})$-plane.
The points on the curves give the stress components realized 
 on the rectangular grids in the $(x,y)$-plane, 
 spaced by $15\Delta{x}$ and $16\Delta{y}$, respectively.
The dotted circle represents the yielding threshold,
$|\bsigma|=\sigmaY$.
(c): The same data as (b) shown on the physical $(x,y)$-plane 
(in units of $\sigmaY$).
(d): The \VIS, $\omega$, for the same data as (b), 
shown by gray scales and contours.
(Only the range of $0\le x \le 60\xi$ is shown in (a,c,d),
 with the axes labeled in units of $\xi$.)
The thick dots ($\bullet$) in (c,d) at 
$ x = 17.0\xi,\, {26.8\xi\;(y>0)},\ {26.8\xi\;(y<0)},\ 
\text{and}\ 46.4\xi $
indicate the  locations where the subsequent yield starts
under the specific \emph{uniform} external shear stresses: 
$(\sigma^{\text{ext}}_{xz},\sigma^{\text{ext}}_{yz})
 = \sigma^{\text{ext}}(\cos\theta,\sin\theta)$ with
$\theta=\pi/2,\; 0,\; \pi,\; -\pi/2$,
respectively  (see the text). 
}
\label{fig:omega}
\end{figure}
%%%%%%%%%%%%%%%%%%%%%%%%%%%%%%%%%%%%%%%%%%%%

%%%%%%%%%%%%%%%%%%%%%%%%%%%%%%%%%%%%%%%%%%%%%%%%%%%%%%%%%%%
To study the characteristics of the temporal evolution of the 
{\VIS}  as well as its influence 
upon the future plastic yielding, we have solved numerically 
Eqs.~(\ref{eq:newton2})--(\ref{eq:wpl2}).
We will focus on the ``hydrodynamic'' or scaling regime whose
spatiotemporal scales are much greater than the viscoelastic
timescale $\tau\equiv \eta/\mu$ and the viscoelastic length-scale
$\xi\equiv {v_{\text{s}}} \tau$.
To this end, we take the system
 of $0\leq x \leq L_x= 120 \xi$  and $|y|\leq L_y/2 = 50\xi$,
and the spatiotemporal mesh sizes,
$(\Delta{x}, \Delta{y}) = (0.390\xi, 0.391\xi) $
and $\Delta{t}=5\tau/16$.
We apply a transient inhomogeneous stress $\sigma_{xz}$ 
on the boundary at $x=0$ 
with 
$ \sigma_{xz} \gtrless 0 $ for $ y \gtrless 0 $
(Fig.~\ref{fig:setup}(a)), of the profile, 
$\sigma_{xz}|_{x=0} = \sigma_{\text{max}} s(t)\, \tanh (y/L_0)$.
Here $\sigma_{\text{max}} = 10\, \sigmaY$, 
$L_0 = 50\, \xi$, and 
 $s(t)$ is a slightly smoothened unit step-function, taking 
the value of unity only during the initial period of about $ 100\tau$, 
with a smoothing period of  approximately $1.5\tau$
(Fig.~\ref{fig:setup}(b)). 
The conditions on the other boundaries have been adjusted to mimic
approximately reflection-free walls~\cite{prep}.
For the temporal evolution
 we have used the two-step Lax-Wendroff (LW) scheme.
The validity of this second-order scheme for the present problem
has been checked in the one-dimensional situation 
by comparing it
 with the Cubic-Interpolated Propagation (CIP) scheme
 of the third-order precision.

%%%%%%%%%%
Fig.~\ref{fig:omega}(a) shows the snapshot of the flux of the \VIS, 
$\bm{J}$,  at $t=100\tau$, the time at which
the external loading is removed. 
The direction of $\bm{J}$ is governed by that of $\bsigma$:
from  (\ref{eq:wpl2}) and (\ref{eq:GS}), we have
$\bm{J}\perp \bsigma$ with $|\bsigma|>\sigmaY$.
%%%%%%%%%%

Figs.~\ref{fig:omega}(b) and \ref{fig:omega}(c) show 
the distribution of the internal (or remanent) stress, 
$\bsigma=\bsigma^{\text{int}}(x,y)$ 
$\equiv (\sigma^{\text{int}}_{xz} (x,y),$ 
$\sigma^{\text{int}}_{yz} (x,y))$,
on the $(\sigma_{xz},\sigma_{yz})$-plane and on the
$(x,y)$-plane, respectively, 
at $t=500\tau$, when all the plastic flow has died out and also
the elastic waves have already left the system through the
reflection-free boundaries 
(therefore $\nabla\cdot \bsigma^{\text{int}}=0$).
Evidently, the amplitude of the internal stress is
by no means stuck to the threshold value, $\sigmaY$,
but is distributed around zero. 
(The maximum value of $|\bsigma|$ in Fig.~\ref{fig:omega}(b,c)
is about $0.515\sigmaY$.)
This is because the stress field is elastically redistributed 
after the plastic deformations so that it finally satisfies
the static balance, $\nabla\cdot \bsigma^{\text{int}}=0$.
In fact it is easy to show, from this condition, that  
the spatial average of $\bsigma$ should be strictly equal 
to zero in the final state with no forces on the boundaries.
%

%%%%%%%%%%
{Fig.~\ref{fig:omega}(d)} shows the distribution of 
the {\VIS}, $\omega$ (see (\ref{eq:IScont})) 
calculated from the data of {Fig.~\ref{fig:omega}(b,c)}.
As compared with the spatial distribution of the internal stress, 
the {\VIS}  reveals a peculiar localized peak,
which is dislocated from 
the center of the noticeable large-scale circular  distribution of the 
stress itself~\cite{barrat}.  
In the present setup of loading,
the {\VIS} ($\omega$) is initially distributed rather broadly 
 in the $y$-direction (not shown),
reflecting the smooth spatial profile of the external loading
  at the boundary of $x=0$.
 However, the flux $\bm{J}$, which drives the {\VIS}
 symmetrically with respect to the $x$-axis 
(see Fig.~\ref{fig:omega}(a)), 
acts to localize the {\VIS} near this axis.
The fan-shaped spatial characteristics of $\omega$ 
in {Fig.~\ref{fig:omega}(d)} suggests an approximate 
self-similarity in the distribution of the \VIS. 
%
%%%%%%%%%%%%%%%%%%%%%%%%%%%%%%%%%%%%%%%%%%%%
\begin{figure}[tb]
 \includegraphics[width=0.90\linewidth] {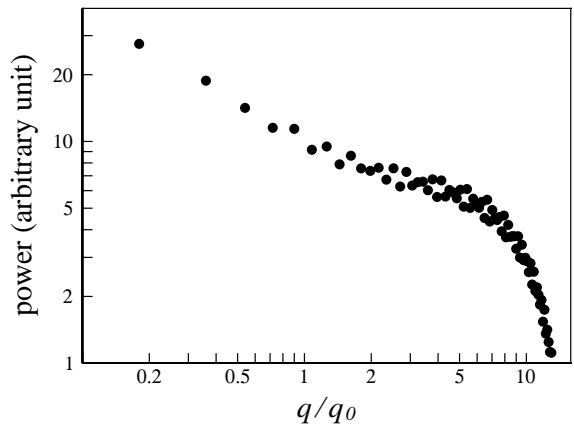}
\caption{The log-log plot of the 
two-dimensional spatial Fourier spectral intensity 
of $\omega$, 
integrated over the angle,
\textit{vs} the amplitude of the wave numbers 
(in units of  $q_0\equiv(2\xi)^{-1}$,
 the wave number beyond which
 the yielded fluid has no propagative modes).
The exponent of the hydrodynamic regime ($q<q_0$) is read out 
to be about $-1/3$.}
\label{fig:spectrum}
\end{figure}
%%%%%%%%%%%%%%%%%%%%%%%%%%%%%%%%%%%%%%%%%%%%
%
This self-similar nature can also be seen in 
the spatial Fourier spectrum of $\omega$ 
(Fig.~\ref{fig:spectrum}), which shows 
a power-law regime in the intermediate range of 
wave numbers.

The state shown in {Fig.~\ref{fig:omega}(b,c,d)}
retains the memory of the plastic flow
 in the past in the form of the {\VIS}.
This memory in turn influences the plastic response to the future
external loading.
As an example, 
if we apply quasi-statically  a uniform external stress 
$\bsigma=\sigma^{\text{ext}}\hat{\bm{t}}(\theta)$,
where 
$\sigma^{\text{ext}}>0$ and 
$\hat{\bm{t}}(\theta)=(\cos\theta,\sin\theta)$,
to the above mentioned state with an internal stress,
the yield then occurs
with a threshold amplitude of $\bsigma^{\text{ext}}$ 
which is generally smaller than $\sigmaY$.
Moreover, both the location $(x,y)$
 and the threshold amplitude of $\bsigma^{\text{ext}}$
at which the yield starts depends on the orientation $\theta$
of the applied uniform stress.
As a demonstration, 
 the thick dots in Fig.~\ref{fig:omega} (c,d) indicate such locations
 for the external shear stress of several
 orientations.
It is interesting to note that the locations where the 
yield starts is well distant from the region where 
{\VIS}  is concentrated.
Moreover, the location of the first yield as a function of $\theta$  
is not necessarily continuous nor single-valued, as seen by 
the two thick dots at $x = 17.0\,\xi $ for the external stress:
$(0, \sigma^{\text{ext}}_{yz})$ with $\sigma^{\text{ext}}_{yz}>0$.

%%%%%%%%%%%%%%%%%%%%%%%%%%%%%%%%%%%%%%%%%%%%%%%%%%%%%%%%

We will describe how the system 
with the internal stress, $\bsigma^{\text{int}}(x,y)$ (see
 Fig.~\ref{fig:omega}(b,c)), 
starts the plastic yield under further quasi-static  
application 
of a {\it uniform} external stress, 
$\bsigma=\sigma^{\text{ext}}\hat{\bm{t}}$. 
First we note that, before the yielding,
the uniform external stress 
is simply superposed onto 
the internal stress, $\bsigma^{\text{int}}(x,y)$, to give  
the actual stress as
$\bsigma^{\text{int}}(x,y) + \sigma^{\text{ext}}\hat{\bm{t}}(\theta)$.
The external stress at the plastic yielding, which we denote by 
$\sigma^{\text{ext}}_{\text{Y}}$, is, therefore, 
the solution of the following equation:
\begin{equation} \label{eq:touch}
\max_{(x,y)}|\bsigma^{\text{int}}(x,y) +
\sigma^{\text{ext}}_{\text{Y}}\hat{\bm{t}}(\theta)| 
={\sigmaY},
\end{equation}
where $\theta$ is fixed.
Here those coordinates $(x,y)$ that realize the maximal value 
on the left hand side tell the location (or locations) 
at which the plastic yielding starts.
The threshold external stress thus obtained plays the role of
apparent yield stress of the
 system with internal stress.
Eq.~(\ref{eq:touch}) is easily interpreted 
and solved graphically: 
As shown in Fig.~\ref{fig:omega}(b), 
the internal stress  $\bsigma^{\text{int}}(x,y)$ on the 
$(\sigma_{xz},\sigma_{yz})$-plane
forms a closed domain.
The superposition of an external uniform stress, 
$\sigma^{\text{ext}}\hat{\bm{t}}(\theta)$, corresponds
to a parallel displacement of this domain by 
$\sigma^{\text{ext}}\hat{\bm{t}}(\theta)$.
If we increase $\sigma^{\text{ext}}$, 
the displaced domain should eventually {touch}
from inside the circle of the yielding condition, 
$|\bsigma|= \sigmaY$ 
(the dotted circle in Fig.~\ref{fig:omega}(b)).

The above graphical solution
immediately leads us to the following three 
qualitative conclusions:
(i) The amplitude of the apparent yield stress 
$\sigma^{\text{ext}}_{\text{Y}}$ is generally 
less than ${\sigmaY}$, the value expected in 
the internal-stress-free system.
(ii)  
If the apparent 
yield shear stress in one direction
is less than $\sigmaY$, then it is true also in the 
opposite direction,
to meet with the requirement of vanishing spatial average of 
the internal stress as mentioned above.
(iii) The yielding point $(x,y)$ 
which solves (\ref{eq:touch}) can be discontinuous 
as the function of the direction of external shear, $\theta$. 
In fact, if we parallelly displace the domain of 
$\bsigma^{\text{int}}(x,y)$ 
by varying $\theta$ while
adjusting the value of $\sigma^{\text{ext}}$ so that 
the domain remains tangent to 
the circle of $|\bsigma|= \sigmaY$, 
the tangential point can undergo jumps.
We expect a similar type of discontinuity to appear 
for the other criteria of plasticity as long as they share 
a similar mathematical structure to (\ref{eq:touch}).

%%%% CONCLUSION %%%%%%%%%%%%%%%%%%%%%%%%%%%%%%%%%%%%%%%%%%%%%%%%%%
In this \textit{Letter}
we have defined and analyzed the internal stress 
 in a continuous model of elasto-plasticity, 
with the emphasis
 on the dynamics of the vorticity of the internal
 stress (\VIS) as the source of the internal stress.
Some of the insights obtained from our analysis 
could be generalized to other rheological models, and 
also to the models of higher dimensionality:
(1) the internal stress after the yielding can be
inhomogeneous,  
(2) the amplitude of the internal stress is generally
less than the threshold value $\sigmaY$, being redistributed
so that the internal balance of elastic force is established,
(3) the apparent threshold yield stress under subsequent global 
shear deformation is different from $\sigmaY$ and anisotropic.
In order to verify and develop these ideas,
it is, firstly,  essential to recognize the quantity
like {\VIS}  which is 
responsible to the internal stress 
and is invariant under elastic deformations.
Secondly, it is important to analyze how such
quantity is created and kept as the memory reflecting the 
past history of the system's plastic flow.
Finally, we should analyze the consequences of 
the inhomogeneous internal stress such as 
the apparent threshold yield stress under
the subsequent global loading.

To be more realistic,
the simple constitutive equation of the {``Bingham type''}
 should
be modified so that it represents the plastic yield as a slow
dynamics involving the effect of temperature and the timescales of
the observation and the operation~\cite{Miyamoto,aradian, Varnik}.
Since the definition of {\VIS} is independent of the inertia effect,
the concept of {\VIS}  should be applicable
to other rheological models in which the momentum can be ignored 
while some internal degrees of freedom of the material are relevant.
Of particular interest would be the relation between 
internal stress and slow relaxation:
the (extended) distribution of relaxation time observed 
through the aging or creep experiments~\cite{paste} could
partly be attributed to the distribution of the internal stress.
In systems consisting of discrete elements, like granular 
materials~\cite{Nakahara}, a macroscopically homogeneous shear will 
be sufficient to cause internal stress among those elements, 
due to 
local anisotropy of elasto-plastic interactions. 
More quantitative analysis on this point 
could be made in future.
As mentioned in the introductory part, the internal stress 
is found in many different systems  not being limited to the 
rheological fluids.
Systems composed of active elements (ex.~biological cells), 
where the internal stress may be generated by the medium itself,
would be highly of interest in this aspect.

%%%%% \begin{acknowledgments} %%%%%%%%%%%%%%%%%%%%%%%%%%%%%%
T.~O.\  
thanks Shin-ichi Sasa and Hiroshige Matsuoka 
for fruitful discussions.  
K.~S.\  
has been partially supported by the Yamada Science Foundation.
The authors also acknowledge A.~Ajdari 
for having drawn our attention to \cite{derec}
during revision of our manuscript.
%%%%% \end{acknowledgments} %%%%%%%%%%%%%%%%%%%%%%%%%%%%%%%%

\end{document}